# Automating Fine Concurrency Control in Object-Oriented Databases

Carmelo Malta and José Martinez

Université des Sciences et Techniques du Languedoc
Laboratoire de Systèmes Informatiques
860, rue de Saint-Priest, 34090 Montpellier, FRANCE
e-mail: {malta, martinez}@crim.crim.fr

**Abstract**

*Several propositions were done to provide adapted concurrency control to object-oriented databases. However, most of these proposals miss the fact that considering solely read and write access modes on instances may lead to less parallelism than in relational databases! This paper cope with that issue, and advantages are numerous: (1) commutativity of methods is determined a priori and automatically by the compiler, without measurable overhead, (2) run-time checking of commutativity is as efficient as for compatibility, (3) inverse operations need not be specified for recovery, (4) this scheme does not preclude more sophisticated approaches, and, last but not least, (5) relational and object-oriented concurrency control schemes with read and write access modes are subsumed under this proposition.*

## 1. Introduction

Several proposals were done to offer concurrency control methods fitted to object-oriented databases. Basically, three propositions consider classical read and write access modes on instances [5, 8, 17], whereas [1] takes into account a finer view of operations by providing instance variable accesses. We argue that considering exclusively read and write accesses on instances is insufficient in an object-oriented database. We must take advantage of commutativity of methods, as several authors did with abstract data types [23, 25]. However, commutativity has the same inherent limitations as compatibility [18]. Then, it is worth conceiving "simple" concurrency control techniques for *arbitrary* objects. The form of commutativity that we introduce is related to the one of [1, 19] but is more conservative. However simple it is, it eliminates four problems which, to our knowledge, have not been addressed in the literature: First, it is unthinkable to put the burden of determining commutativity of *every pair* of methods (and providing inverse operations, for recovery, of *every* method) on the application programmer. Also, code reuse leads to locking overhead, several lock escalations and deadlocks. Lastly, with read and write access modes alone, *unreasonable* conflicts occur because several of these do *not* appear in relational databases!

The organization of the paper is as follows: First, we introduce the basic concepts of object-oriented databases, relevant to a number of currently implemented systems. Then, we detail the four major problems which make read and write accesses to instances unsatisfactory. Next, section 4 defines direct access vectors and provides the outline of an efficient algorithm to compute transitive access vectors, the solution to the mentioned problems. In section 5, we describe the use of transitive access vectors in the locking protocol of an inheritance graph. Lastly, we compare our work to previous ones. Section 7 concludes this paper.

## 2. An object-oriented database

To be useful for a majority of object-oriented systems, we consider the highest common factor of object-oriented data models. We shall insist on the calling mechanism which brings into play inheritance, overriding, and late binding.

### 2.1. Data model

The most commonly described object-oriented data model is class-based. It distinguishes instances and classes, (but not meta-classes.) Instances pertain to exactly one class. Classes are related by simple or multiple inheritance. These are the basic concepts introduced by Smalltalk [9], which can be found in ORION [3], $O_2$ [16], GemStone [4], ObjectStore [15], or VBASE [2]. Neither IRIS [7] nor G-BASE [22] are directly concerned since the former allows multiple

This work was supported in part by the PRCs BD3 and C3 coordinated by the Centre National de la Recherche Scientifique (CNRS), and in part by the Ministère de la Recherche et de la Technologie (MRT).

instanciation, whereas the latter deals with meta-classes.

A class is composed of a tuple describing instance variables, which we shortly call fields, and of a set of methods, only way to manipulate instances. We differentiate fields which are base types, such as integers or characters, from those which reference other instances (e. g., instances of class "Person" may have a field "Father" which references another instance of the same domain.) Some databases, as $O_2$, offer complex types, i. e., fields may be tuples, sets, bags, lists, but we do not investigate that issue [11].

Fields and methods are inherited by a subclass from its superclasses. The subclass can add new fields and new methods to its definition. Also, the code of an inherited method can be overridden in a subclass. This feature complicates our technique and deserves further presentation.

## 2.2. Calling methods

The message paradigm of object-oriented programming consists in sending messages to objects rather than applying procedures or manipulating them directly. This paradigm achieves *encapsulation*. A message is linked at run-time to a method, depending on the class of the instance. This is the *late binding* mechanism. The receiver is particularized in the code of the method; when explicitly required, it is generally named **self**.

In the code of a method, we are not at all interested in control structures. Then, the code is abstracted as a sequence of assignments, expressions and messages. Messages are further divided into two subcases: simple and prefixed messages.

The first form is the more usual. Self-directed messages are linked to the more appropriate method, e. g., one which is located in the nearest ancestor class of the instance class. We use the syntactic form "**send** M **to** f" where M is a method name and f the instance (variable) on which M is requested.

The prefixed form is to be found when a method is redefined not completely, but as an extension of the replaced one. Then, the code of the overriding method contains a call to the overridden one. We note it with "**send** C.M **to** f" where C is an ancestor class of the proper class of f from where the method M is to be taken.

## 3. The four problems

In this section, we highlight four problems related to concurrency control in object-oriented databases, namely difficulty to provide *ad hoc* commutativity relations, locking overhead, lock escalation, and pseudo-conflicts.

*commutativity of methods*

Several abstract data types (ADTs) can be implemented once. It is of interest to assign to their operations fine and *ad hoc* commutativity relations [21]. Examples of such ADTs are sets, maps, stacks, counters, etc [23, 25].

Classes have a direct relationship with ADTs but they also have a meaningful difference: They are related to each other by the inheritance relationship. Therefore, two methods with the same name may have distinct properties of commutativity. Also, modifying a method in a given class may modify several of its subclasses.

From this observation, we conclude that automation of the determination of commutativity is primordial in object-oriented databases when methods are frequently added, removed, or updated. Moreover, it is unthinkable to put the burden of determining commutativity of *every pair* of methods (and providing inverse operations, for recovery, of *every* method) on the application programmer.

Note that we do not discard the use of *ad hoc* commutativity relations. It is of interest for predefined types or classes, as the "Integer" type or the "Collection" class, to be delivered with high commutativity performances (See, for example, [20].)

We introduce the example of Figure 1 to illustrate the three other problems. It represents a rather simple but instructive hierarchy. We note that fields are either of a predefined type, as integer or boolean, or reference instances of other classes, as $f_3$.

We then turn our attention to methods. Some of them are quite general, as $m_1$ in $c_1$; implementation details are deferred to other methods, $m_2$ and $m_3$, which may be overridden in subclasses, as $m_2$ in $c_2$, to specialize the algorithm. Sometimes, a method is overridden not completely but as an extension of the inherited version; it is the case for $m_2$ in $c_2$. All this is *code reuse*. This kind of object-oriented programming is powerful but leads to a great number of self-directed messages. A last remark about methods is that they appear arbitrarily at different levels in the hierarchy, as $m_4$ which is not defined in $c_1$.

All this seems trivial, but examining it from the concurrency point of view shows that propositions which only recognize read and write access modes are insufficient. In that case, $m_1$ and $m_3$ are readers while $m_2$ is a writer in class $c_1$. In class $c_2$, $m_1$ and $m_3$ cannot change since they are inherited, and $m_2$ and $m_4$ are writers. We have three problems:

   (i) one instance can be controlled several times for what can be considered one actual access;
   (ii) deadlocks can occur due to lock escalation;
   (iii) two methods, both classified as writer, but manipulating different fields, conflict unnecessarily.

*locking overhead*

Code reuse leads to self-directed messages. Since it is a powerful programming technique, one should expect it to be often used in an object-oriented database. This must not become a bottleneck for object-oriented concurrency control. If each message wants control, then invoking $m_1$ on an instance of $c_1$ or $c_2$ leads to controlling concurrency *thrice*.

*lock escalation and deadlocks*

System R measurements are often cited [14]: 97 % of deadlocks are due to lock escalation from read to write mode; up to 76 % of these deadlocks are avoided by announcing the more exclusive access mode.

It is exactly what happens with $m_1$: it acquires a read lock on a given instance, then, the message $m_2$ is sent which requires, (at least for instances of $c_1$ and $c_2$), a write lock. This write lock could have been requested immediately when $m_1$ was sent, which would have eliminated some risks of deadlocks.

*pseudo-conflicts*

It is reasonable to expect a method which has been classified as a writer (respectively a reader) to find its overriding counterparts classified in the same set. It is true for method $m_2$. But, new methods appear in subclasses, some of which manipulate exclusively added fields. The dichotomy between readers and writers does not take into account this reality. It is the case with $m_2$ and $m_4$ in class $c_2$: $m_4$ does not exist in $c_1$ and it accesses only to fields defined in $c_2$. Nevertheless, $m_2$ and $m_4$ conflict, which is *unreasonable*!

In point of fact, this problem does *not* appear in relational databases. Let us represent $c_1$ and $c_2$ in a relational schema: they become respectively relations $r_1$ and $r_2$. Assuming that field $f_1$ is the primary key of $r_1$, $r_2$ contains the fields defined in class $c_2$ plus $f_1$ as a foreign key. If a transaction accesses to all the fields of class $c_2$, then a join operation is executed, and $r_1$ and $r_2$ are submitted to concurrency control. By contrast, if a transaction just accesses to the fields defined in $c_2$, then only $r_2$ will be locked, therefore allowing a concurrent transaction to access to $r_1$. This concurrent execution is obtained in a relational database without considering a smaller granule of locking than the tuple. In object-oriented databases, it is *mandatory* to take a smaller granule of locking than the instance.

In this section, we have highligthed four problems which have not been addressed in the literature. The last one is certainly the most important. Our proposition eliminates these four problems thanks to a quite simple analysis of the source code of methods at compile-time and an efficient algorithm based on determination of strong components in directed graphs. One access mode per method per class is generated and used as a conventional access mode [13]. Consequently, no performance penalty is incurred at run-time.

Roughly speaking, the technique consists in associating to each method a *direct access vector*. A class is a cartesian product of the domains of its different fields. To each method in each class where it is defined, we associate a vector of the same dimension as the cartesian product. Each value composing this vector will denote the most restrictive access mode used by the method when accessing to the corresponding field. Commutativity of methods is determined by comparing access vectors. (Recovery uses access vectors as projection patterns for extracting the modified parts of instances, but it is not discussed here.)

Direct access vectors eliminate the first and fourth inconveniences. To eliminate the second and third ones, *transitive* access vectors have to be constructed.

## 4. Constructing access vectors

The task of computing transitive access vectors is not completely obvious even when solely ADTs are involved because methods can call each other recursively. Inheritance further complicates the situation: we have to deal with *multiple inheritance*, *overriding*, and *late binding*. Therefore, access vectors are defined in a less straightforward way than informally

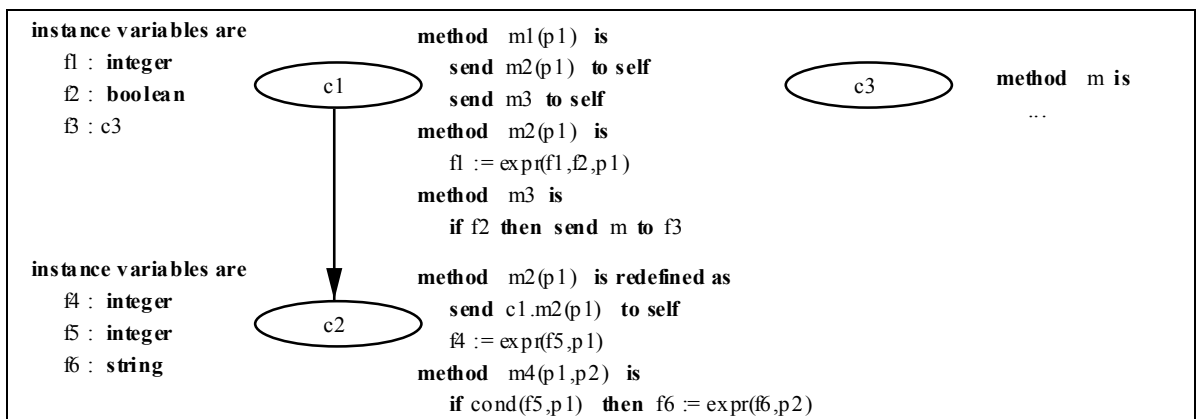

**Figure 1**: An example of object-oriented programming

introduced.

## 4.1. Preliminary definitions

First of all, the notion of source code of a method is not formalized since we just need to detect field assignments, expressions and messages sent to the current instance itself. Therefore, we rely on an informal meaning of source code.

*definition 1*
Each class is a pair composed of a set of fields and of a set of methods, respectively denoted FIELDS(C) and METHODS(C) for a given class C. The notation FIELDS(a) is also used with access vectors. Also, we note ANCESTORS(C) the set of classes from which C inherits, directly or transitively.

*definition 2*
We call $c_{MODES}$ the binary compatibility relation on MODES, given in extension in Table 1, where MODES = {Null, Read, Write} with Null < Read < Write.

|       | Null | Read | Write |
|-------|------|------|-------|
| Null  | yes  | yes  | yes   |
| Read  | yes  | yes  | no    |
| Write | yes  | no   | no    |

**Table 1**: Classical compatibility relation

The order relation on MODES is directly deduced from the compatibility relation by inclusion of rows and columns [13]. We will use the join operator ($\Delta$) of the theory of lattices on MODES. (On a total order, join is equivalent to max, e. g., Read $\Delta$ Write = Write.)

*definition 3*
An *access vector* for a method M in a class C is a bag of modes indexed by the fields of C:
$AV_{C,M} = (m_f \in MODES)_{f \in FIELDS(C)}$

For example, the direct access vector of $m_2$ in $c_1$ is (Write$_{f_1}$, Read$_{f_2}$, Null$_{f_3}$).

We extend the join operator on MODES to access vectors.

*definition 4*
Let a' et a" be access vectors, the join operator on them is defined over:
$(m_f \in MODES)_{f \in FIELDS(a') \cup FIELDS(a")}$
such that:
a' $\vee$ a" = $(m'_f \vee m''_f)_{f \in FIELDS(a') \cap FIELDS(a")} \cup$
$(m'_f)_{f \in FIELDS(a') \setminus FIELDS(a")} \cup$
$(m''_f)_{f \in FIELDS(a") \setminus FIELDS(a')}$

Calculating the join of two access vectors is collecting all the fields and taking the most restrictive access mode for common fields. For example, (Write$_X$, Read$_Y$, Read$_Z$) $\vee$ (Read$_X$, Null$_Y$, Read$_T$) = (Write$_X$, Read$_Y$, Read$_Z$, Read$_T$).

The algorithm of subsection 4.3 requires the following straightforward property.

*property 1*
The join operator on access vectors is idempotent, commutative, and associative.

We end off this subsection with the unsurprising definition of commutativity of access vectors.

*definition 5*
Let a' and a" be access vectors, then we note **c** the commutativity relation given by:
a' **c** a" $\diagup$
$\forall$ f $\in$ FIELDS(a') $\cap$ FIELDS(a"), m'$_f$ **c**$_{MODES}$ m"$_f$

## 4.2. Compiling methods

To determine commutativity of methods, their source codes are parsed. In this subsection, we give three definitions which are the specifications of the information which must be extracted by the compiler from any method. Definition 6 gives the direct access vectors of a method. The sets of definitions 7 and 8 serve to construct, in the following subsection, the late binding resolution graph of each class, a prerequisite for calculating transitive access vectors.

*definition 6*
Let C be a class, then to each method M in METHODS(C), we associate a *direct access vector*, DAV$_{C,M}$, such that:
(i) if M is inherited from a superclass C', then:
DAV$_{C,M}$ = DAV$_{C',M}$ $\vee$ (Null$_f$)$_{f \in FIELDS(C)}$
(ii) otherwise, if M is defined for the first time or overridden in C, then:
$\forall$ f $\in$ FIELDS(C),
 Write$_f$ $\in$ DAV$_{C,M}$ $\Leftrightarrow$ there is an assignment of the form "f := <expression>" in the code of M;
 Read$_f$ $\in$ DAV$_{C,M}$ $\Leftrightarrow$ there is no such assignment, but "f" appears in some expression, including messages;
 Null$_f$ $\in$ DAV$_{C,M}$ $\Leftrightarrow$ "f" appears nowhere in M.

In point of fact, distributing the fields of an instance over several relations, as done in section 3, and then locking the relations separately, is creating a coarse access vector: When answering to a request, each accessed relation will be locked either exclusively, or in shared mode, whereas unuseful relations are not accessed, i. e., "Null-locked."

*definition 7*
Let C be a class, then to each method M in

METHODS(C), we associate a set of *direct self-calls*, $DSC_{C,M}$, defined as follows:

(i) if M is inherited from C', then:
$DSC_{C,M} = DSC_{C',M}$

(ii) otherwise:
$DSC_{C,M}$ = { M' $\in$ METHODS(C) | the message "**send** M' **to self**" appears in the code of M }

It is the sets of direct self-calls which solve, *at compile-time*, late bindings which occur *at run-time*! The reader is requested to wait until the following subsection.

*definition 8*

Let C be a class, then to each method M in METHODS(C), we associate a set of *prefixed self-calls*, $PSC_{C,M}$, defined as follows:

(i) if M is inherited from C':
$PSC_{C,M} = PSC_{C',M}$

(ii) otherwise:
$PSC_{C,M}$ = { (C',M') | C' $\in$ ANCESTORS(C), M' $\in$ METHODS(C'), and the message "**send** C'.M' **to self**" appears in the code of M }

Note how simple it is, for a compiler, to construct the direct access vector (DAV) as well as the direct (DSC) and prefixed self-calls (PSC) sets of any method.

Direct access vectors of definition 6 may be sufficient for ADTs, but with object-oriented programming we cannot just rely on them. As mentioned in section 3, it is worth controlling concurrency only once per instance, i. e., solely when the top message is sent. Hence, we have to construct *transitive* access vectors.

### 4.3. An algorithm

In this subsection we just give the parameter of the algorithm because its core is the well-known problem of determining strong components of a directed graph, which has been solved efficiently a long time ago [24].

From the informations extracted after parsing the codes of all the methods of a class C and of its ancestors, we construct the *late binding resolution graph* which is applicable to any proper instance of C.

*definition 9*

Let C be a class, then $G_C(V,\Gamma)$ is its *late binding resolution graph*, where:

$V$ = { {C} $\times$ METHODS(C) } $\cup$

$\bigcup_{M \in METHODS(C)} PSC^*_{C,M}$

$\forall$ (C',M') $\in$ V,

$\Gamma$(C',M') = { {C} $\times$ $DSC_{C',M'}$ } $\cup$ $PSC_{C',M'}$

where $PSC^*$ is the reflexo-transitive closure of PSC.

To illustrate this awkward definition, let us construct the late binding resolution graph of class $c_2$. Each set of prefixed self-calls is empty except $PSC_{c_2,m_2}$ which is equal to {$(c_1,m_2)$}, and to its reflexo-transitive closure too since $PSC_{c_1,m_2}$ is empty. Hence, V is equal to {$(c_2,m_1)$, $(c_2,m_2)$, $(c_2,m_3)$, $(c_2,m_4)$} $\cup$ {$(c_1,m_2)$}. Also, the sets of direct self-calls are all empty but $DSC_{c_2,m_1}$ which is equal to {$m_2$, $m_3$}. It serves to construct the edges (($c_2,m_1$), ($c_2,m_2$)) and (($c_2,m_1$), ($c_2,m_3$)), while the edge (($c_2,m_2$), ($c_1,m_2$)) is given directly by $PSC_{c_2,m_2}$. Figure 2 is the resulting graph.

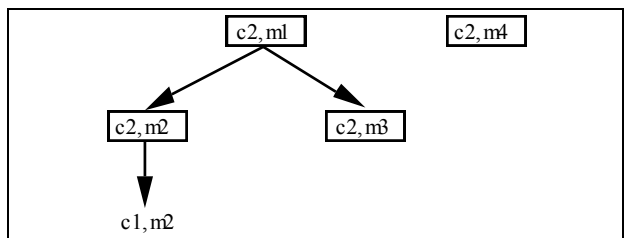

**Figure 2**: The late binding resolution graph of class $c_2$

Transitive access vectors are easily computable from this graph.

*definition 10*

Let C be a class and M a method defined in C, then we define the *transitive access vector*, $TAV_{C,M}$, as follows:

$TAV_{C,M} = DAV_{C,M} \vee \bigvee_{(C',M') \in \Gamma^*(C, M)} DAV_{C',M'}$

where $\Gamma^*$ is the reflexo-transitive closure of $\Gamma$.

The definition of the value of the transitive access vector of a method M in a class C is quite simple: it is the join of the direct access vectors of all the methods which may be executed when M is sent to a proper instance of C. The algorithm is a depth-first search whose time complexity is in $O(|V| + |\Gamma|)$, i. e., *linear* in the size of the graph. Transitive access vectors are calculated from the sinks, with the obvious equality between TAV and DAV, up to the sources.

With the graph of Figure 2, the transitive access vectors of ($c_1,m_2$), ($c_2,m_3$), and ($c_2,m_4$) are equal to their respective direct access vectors, i. e., ($Write_{f_1}$, $Read_{f_2}$, $Null_{f_3}$), ($Null_{f_1}$, $Read_{f_2}$, $Read_{f_3}$, $Null_{f_4}$, $Null_{f_5}$, $Null_{f_6}$), and ($Null_{f_1}$, $Null_{f_2}$, $Null_{f_3}$, $Null_{f_4}$, $Read_{f_5}$, $Write_{f_6}$). Then, the transitive access vector of ($c_2,m_2$) is the join of its direct access vector, ($Null_{f_1}$, $Null_{f_2}$, $Null_{f_3}$, $Write_{f_4}$, $Read_{f_5}$, $Null_{f_6}$), and of the transitive access vector of $(c_1,m_2)$, which gives $(Write_{f_1}, Read_{f_2}, Null_{f_3}, Write_{f_4}, Read_{f_5}, Null_{f_6})$. At last, $TAV_{c_2,m_1}$ is computed from $TAV_{c_2,m_2}$ and $TAV_{c_2,m_3}$, giving $(Write_{f_1}, Read_{f_2}, Read_{f_3}, Write_{f_4}, Read_{f_5}, Null_{f_6})$.

Since a method can call itself recursively through other methods, directed cycles can appear in these graphs. We make the obvious observation that transitive access vectors of vertices pertaining to a common directed cycle are necessarily equal since their respective $\Gamma^*$ are identical. Hence, we can still calculate transitive access vectors with a single depth-first search by using the algorithm of [24] for determining strong components. Thanks to property 1, cyclic dependencies are computable (idempotence) in any order (commutativity and associativity.)

One might object that such graphs are perhaps unbearable to manage. We just remark that a more complex graph is proposed in $O_2$ [26]. Not solely classes related by inheritance are concerned but also classes related by composition; this huge structure is called the *method dependency graph*. Thus, our proposition can be merged elegantly with previous works.

In this section we achieved to deliver a distinct access mode to each method, rather than just classifying it as a reader or a writer. Besides, we showed that there exists an efficient algorithm to calculate transitive access vectors which take into account not solely the code of a method but its whole pattern of execution on the current instance, **self**. This decreases significantly the number of controls.

A disadvantage of transitive access vectors is that they are very conservative. They even represent impossible executions because they forget alternatives in the analysis of the source codes. This is not so much a problem when, and it is often the case in a database, methods are applied to sets of instances, because each pattern of execution of the method is probable. Furthermore, hierarchical locking [10] would have been impossible otherwise.

## 5. Locking in an inheritance graph

[8] and [17] elaborated locking protocols on inheritance graphs which can lock implicitly some classes. This was possible only because access modes on instances were mere reads and writes and, consequently, characterized any method in any class. Now, we have an access mode per method per class, and, consequently, they are no longer defined on any class. Thus, we have to rely on explicit locking of classes. (Note that this justifies, *a posteriori*, the "somewhat arbitrary" (sic) choice made for ORION [12].) For locking to be cheap, access vectors will be first translated into access modes.

### 5.1. From access vectors to access modes

Using transitive access vectors as locks leads to an overhead, compared to read and write locks, due to their length. To eliminate this drawback, access vectors are translated into access modes. One commutativity relation per class is created; an access mode per method is produced. Two access modes commute if, and only if, their respective transitive access vectors commute according to definition 5. The commutativity relation of class $c_2$ is given in Table 2. (Commutativity relation of class $c_1$ is obtained, in this example, as the restriction of Table 2 to $m_1$, $m_2$, and $m_3$.) From the principle of construction of access modes, we know that the parallelism which is allowed by access modes is exactly the one which is permitted by access vectors.

|       | $m_1$ | $m_2$ | $m_3$ | $m_4$ |
|-------|-------|-------|-------|-------|
| $m_1$ | no    | no    | yes   | yes   |
| $m_2$ | no    | no    | yes   | yes   |
| $m_3$ | yes   | yes   | yes   | yes   |
| $m_4$ | yes   | yes   | yes   | no    |

**Table 2**: Commutativity relation of class $c_2$

### 5.2. The locking protocol

We rely on strict two-phase locking [6]. Accesses to instances can be classified as:
  (i) accesses to one instance of one class;
  (ii) accesses to a majority of instances, if not all, of one class;
  (iii) accesses to some instances of all the classes rooted at C, i. e., pertaining to a common domain;
  (iv) accesses to a majority or all the instances of all the classes of domain C.

Since, at the class level, implicit locking is no longer feasible, locking an individual class or all the classes belonging to the same domain is essentially the same. Nonetheless, at the instance level, implicit locking is still useful. If a transaction accesses to all the instances of a class, then it is worth locking uniquely the class in hierarchical mode instead of each instance individually. Therefore, an access mode is also a lock on instances, but a lock on a class is a pair composed of an access mode and of a boolean indicating whether locking is hierarchical (as S and IS, X and IX in [10].)

*access to one instance*
  When transaction $T_1$ sends the message $m_1$ to an instance i of $c_1$, the lock $m_1$ is acquired on i, and the lock $(m_1,false)$ on $c_1$.

*access to all instances of a class*
  When the message $m_1$ is sent by transaction $T_2$ to the extension of class $c_1$, no lock is acquired on any instance, but the lock $(m_1,true)$ is requested on $c_1$ and $c_2$. As the lock held by $T_1$ is intentional while the one asked for by $T_2$ is hierarchical, commutativity depends on the access modes which are incompatible (See Table 2.) Hence, $T_2$ and $T_1$ cannot be concurrent.

*access to some instances of a domain*

Another transaction, $T_3$, is sending the message $m_3$ to several instances of the domain rooted at $c_1$. Then, classes $c_1$, $c_2$, and other subclasses of $c_1$, are locked with $(m_3,false)$. Each actually used instance will be locked with $m_3$. $T_3$ can run concurrently either with $T_1$ (if they do not access to common instances), or with $T_2$.

*access to all instances of a domain*

A last transaction, $T_4$, wants to send $m_4$ to all instances of the domain rooted at $c_2$. The lock $(m_4,true)$ has to be acquired on every classes of domain $c_2$. Neither of the preceding transactions can block $T_4$.

Therefore, thanks to transitive access vectors, either $T_1\|T_3\|T_4$, or $T_2\|T_3\|T_4$ are allowed.

With read and write access modes alone, either $T_1\|T_3$ would have been allowed since both use intentional locking, or $T_1\|T_4$ because they do not share any instance.

In the associated relational schema (See section 3), $T_1$ locks one tuple of $r_1$ in write mode and the associated tuple of $r_2$ in write mode too (because $f_1$, the primary and foreign key is modified), $T_2$ locks both relations in write mode, $T_3$ locks $r_1$ in read mode, and $T_4$ locks $r_2$ in write mode. Consequently, either $T_1\|T_3$, or $T_3\|T_4$ are allowed.

Note that permitted concurrent executions are incomparable. In point of fact, the kind of separations which are achieved by inheritance and first normal form are orthogonal: the former offers a kind of predicative locking, and the latter a rough form of field locking. Both previous concurrency control schemes are subsumed whithin our framework.

Another important remark is that $T_1\|T_3\|T_4$ (but not $T_2\|T_3\|T_4$) would have been allowed in the relational schema if $m_2$ did not modify the key field. This is why object-oriented databases implemented on top of relational databases, like IRIS [7], do not feel the need for a special concurrency control method, because object identifiers (OIDs) play the role of primary and foreign keys.

## 6. Related works

Historically, access vectors were already proposed by [6] in conjunction with predicative locking. In System R, predicative locking was abandoned and field locking alone was no longer considered. Some reasons may be that (1) it is expensive to parse each SQL request, at run-time, to construct access vectors, (2) it is also expensive to lock with access vectors of varying length, and (3), as seen in sections 3 and 4.2, first normal form decomposition looks like coarse access vectors.

We saw in section 3 that we need access vectors to obtain some parallelism which occurs in relational databases. Thanks to the fact that classes encapsulate both data and methods, access vectors are determined at compile-time. At last, access vectors are translated into mere access modes, hence this method does not incur locking overhead at run-time. None of the problems mentioned above remains in object-oriented databases.

Very recently, [1] proposed field locking. Basically, the method consists in associating to each class two set ADTs: one for the methods, one for the fields (according to definition 1.) When a message is sent, the activated method is locked in the method set ADT. Then, each field accessed by this method must be locked in the field set ADT. Obviously, this technique achieves field granularity locking. As field locking is done individually at run-time, this technique incurs a much higher overhead. Also, the problems of multiple controls and deadlocks due to escalation are not resolved. In counterpart, this approach is less conservative than ours.

We think that the choice between this technique and ours depends on the frequency of updates. For continuously evolving schemas, the framework of [1] is largely preferable (though schema evolutions are quite exclusive operations.) For applications which do not change perpetually but solely at regular intervals of time, ours is to be chosen. In point of fact, it is the same as choosing between an interpreter (e. g., ORION and Lisp) and a compiler (e. g., $O_2$ and C.)

## 7. Conclusion

In this paper we have highlighted four important problems which render some previous propositions less effective. The most important of all is that some parallelism is lost in object-oriented databases with respect to relational ones. All of these problems can be solved by providing a simple form of commutativity. This kind of commutativity is syntactically extracted from the source codes of the methods at compile-time. Then, an efficient (linear) algorithm calculate what we called transitive access vectors. Finally, transitive access vectors are translated into classical access modes in order not to incur performance penalty at run-time.

What makes this proposition so attractive is that the whole technique is easily implementable and efficient.

This is a major advantage in the field of object-oriented databases when methods are expected to be regularly created, deleted, or updated. At last, finer techniques are not discarded of our framework.